\documentclass[conference]{IEEEtran}
\IEEEoverridecommandlockouts

\usepackage{cite}
\usepackage{amsmath,amssymb,amsfonts}
\usepackage{algorithmic}
\usepackage{graphicx}
\usepackage{textcomp}
\usepackage{xcolor}
\def\BibTeX{{\rm B\kern-.05em{\sc i\kern-.025em b}\kern-.08em
    T\kern-.1667em\lower.7ex\hbox{E}\kern-.125emX}}

\usepackage{hyperref}

\newlength{\halfwidth}
\begin{document}

\title{Implicit Test Oracles for Quantum Computing}

\author{\IEEEauthorblockN{\href{http://www.cs.ucl.ac.uk/staff/W.Langdon}
{William B. Langdon}}
\IEEEauthorblockA{\textit{Department of Computer Science} \\
\textit{University College London,
Gower Street}\\
London, UK \\
w.langdon@cs.ucl.ac.uk}
}

\maketitle

\begin{abstract}
Testing can be key to software quality assurance.
Automated verification may increase throughput
and reduce human fallibility errors.
Test scripts supply inputs, run programs and check their outputs
mechanically using test oracles.
In software engineering 
implicit oracles automatically check for universally undesirable
behaviour, such as the software under test crashing.
We propose 4 properties 
(probability distributions,
fixed qubit width,
reversibility
and
entropy conservation)
which all quantum computing must have
and suggest they could be implicit test oracles for
automatic, random, or fuzz testing of quantum circuits and
simulators of quantum programs.
\end{abstract}


\section{Introduction}

\noindent
Testing is one of the main ways of verifying that software is
sound and does what it should.
Conventionally this means running it on one or more test cases
and comparing its output with the desired output.
Which means that the test engineer has to both specify
the inputs to the program and its required behaviour,
typically as a set of inputs and outputs per test.
A~{\em test oracle} is used to check the program's behaviour.
E.g.~the oracle checks 
that the program's output is as expected.

If the program gives an acceptable answer for a test input
then the test oracle says the program has passed the test.
If not, the oracle says the software under test has failed.
An automatic test script may run the program on 
each of the available tests in the test suite
and count how many pass and how many fail
or it may be set to stop the first time the program
fails a test.
The  test suite may contain many test cases
but often there is no formal specification
from which they could be derived.
Intensive testing can be effective,
however it needs many tests 
but it can be too labour intensive
for a developer to write them by hand.

It might be sufficient to generate test cases at random.
If not there are increasingly sophisticated tools,
including fuzz testing,
for generating test inputs which exercise more
of the program.
Fuzzing in particular
has proved to be very successful at finding bugs,
even in mature software, by prolonged exercising of the code.
Fuzzing campaigns can be measured in hours or even days.
However
often there is no easy way of knowing in advance what 
output a program should give on a mechanically generated input.

To avoid having to devise by hand a large number of test cases
with associated requirements on their answers,
automated testing, particular fuzz testing uses
{\em implicit test oracles}.
I.e., general automatic ways to recognise that the software under test has
misbehaved.
Typical implicit oracles are:
the program should not crash,
e.g.\ it should not cause a {\em null pointer exception} (NPE)
or {\em segmentation error} (segfault),
and it should not {\em loop infinitely}.
Typically an implicit test oracle for non-terminating programs
imposes a program specific timeout before which it should stop running.
That is,
an implicit oracle is an automatic way of recognising software
has failed 
which holds for almost all software.

In the next section we proposed four
{implicit test oracles for quantum circuits}
and simulators of quantum programs
which exploit properties all correct quantum programs must have.
Therefore, while they do not ensure that a quantum circuit is correct,
if they fail, 
we can be confident that something is wrong somewhere.
Section~\ref{sec:discussion}
discusses the difficulties of converting 
these properties into implicit test oracles,
which may be useful for both random and quantum fuzz testing.
\label{p.wang}
(Note Wang et al.'s QuanFuzz
\cite{wang2018quanfuzzfuzztestingquantum,Wang:2021:ICST}
ensures the correct answer is available,
whilst
Blackwell et al.~\cite{Blackwell_2024_SSBSEChallenge}
use differential fuzzing, i.e.~compare results from different simulators,
rather than using implicit oracles.)

\section{Implicit Quantum Test Oracles}
\label{sec:qoracles}

\subsection{Measurement is Probabilistic}
\label{sec:pdistribution}

\noindent
At some point all quantum computation requires measurements
to be taken.
Measurement of the quantum wave form collapses it from a superposition
of states to a definite (classical) state.
Which state is determined by a probability distribution.
Well designed quantum circuits 
will ensure that the state corresponding to desired answer is
highly probably.
However in practise, by repeated operation,
it may be sufficient that acceptable answers are simply more
likely than others.
Nonetheless,
both in simulation and in real quantum circuits
a probability distribution must have the following properties:

\begin{itemize}
\item probabilities cannot be negative or exceed 1.0
\item probabilities must sum up to 1.0
\end{itemize}
If either property fails,
then we can be confident that there is an error somewhere.
E.g.\ counting the states.

\subsection{The Number of Q-Bits is Conserved}
\label{sec:numq}

\noindent
Although a quantum circuit can contain many different quantum gates
connected its Q-bits in many ways,
at all times the width (the total number of Q-bits)
must be the same.
For example,
if the input accepts three Q-bits
then the circuit must always be 3 Q-bits wide
and it must generate three Q-bits of output.

\subsection{Components of Quantum Circuits are Always Reversible}
\label{sec:reversibilty}

\noindent
Unlike classical computing,
quantum circuits are reversible.
Meaning
if any given (test) superposition of states is injected
into any quantum circuit and gives 
rise to a superposition at the output (result),
then feeding the result superposition into the output
will give again give rise to the test superposition at the input.

Further, this is true of any fragment of the circuit,
including individual quantum gates.

\subsection{Quantum Reversibility Means No Entropy Change}
\label{sec:entropy}

\noindent
Shannon entropy is a useful measure of information content~%
\cite{Shannon.Weaver:64}.
The entropy of a probability distribution is
$- \sum p_i \log_2( p_i)$.
Where the sum is over the whole distribution and $p_i$
is the probability of event~$i$.
Using $\log_2$ (logarithm to the base~2)
means information content is expressed in terms of number of bits.
The use of product and logarithms
means entropy is a real number and can give
non integer results.

Excluding measurements, 
in quantum circuits
no information is lost.
Which means the entropy of a quantum (reversible) circuit's input
is the same as the entropy of its output.

This is very different from normal (irreversible) computing,
where instead of having a one-to-one mapping of input-to-output
it is very common that a program can generate
the same output for many inputs.
E.g.\ the integer divide-by-two program
outputs 1 for both input 2 and 3.
Thus for a uniform distribution of inputs from 0 to 3,
the input entropy is $-\sum 1/4 \times \log_{2}{1/4}$
= 2 bits.
But there are only two possible outputs (0 and 1)
and they are equally likely
(entropy~$=-\sum 1/2 \times \log_{2}{1/2}$ = 1 bit).
Therefore the program has lost entropy
(input $2.0$~bits $\Rightarrow$ output $1.0$~bits).

In non-quantum programs,
it is very common for many inputs to give the same answer.
(They are $n$-to-1, where $n$ can be huge).
E.g., given random text as input a program to check spelling will
often say the text is not valid english.
That is, there are a huge number of inputs 
which producing the same  ``not valid'' answer.
Supposing the random text contains $m$ words which are equally likely,
the entropy of the input will be $\log_{2}m$~bits
but 
if the program only produces either ``valid'' or ``not~valid'' answers.
The entropy of the output is at most one bit.
Meaning almost all the entropy of the input has been lost.
This is typical of ordinary everyday mundane computing.

Notice with quantum computing
we have to be careful how we calculate entropy,
particularly whether we include measurement or not.
Once measurements are taken the circuit is not reversible
and the traditional entropy calculations (above) can be used
and the implicit entropy test oracle must report an error
if entropy increases

Entropy is a property of distributions
and for the quantum part of the circuit,
we must be clear that entropy is taken over the distribution
of distributions of entangled states.
I.e.\ entropy is calculated over the
distribution of probability distributions.
Since a quantum circuit is reversible it
must be $n$-to-$n$,
so there must be as many distributions of 
quantum superpositions at its output as there are at its input.
Entropy potentially gives an easy way of applying a sanity check
on what could quickly turn into an exponentially large
number of different probability distributions.
Note for the quantum circuits the {\em implicit entropy test oracle}
must report an error if entropy either increases or decreases.

We can think of entropy as measuring information content.
Reversible (quantum) computing does not loose information,
therefore there can be no lost of entropy.
So if entropy has changed, this means there is an error somewhere
and a quantum fuzzer could potentially 
use this as another way to automatically flag errors.

\section{Discussion: Using Implicit Q-Oracles}
\label{sec:discussion}

\noindent
In some domains, 
particularity discovering potentially security issues,
fuzz testing of system utilities written in high level languages,
has been very successful.
Fuzz testing initially blindly exercises the software under test
but unlike simple random testing,
it is designed to exploit knowledge of the SUT's source
code to create new tests that run parts of the code that have
not been used much so far.
Since the fuzzer knows only the test inputs,
it cannot know the correct answer
but instead relies of implicit test oracles,
such as the program failed 
(e.g.~crashed with a non-existent memory error, NPE)
to flag interesting problems to the test engineer.

Fuzz testing has already be proposed for
testing quantum simulators
(Section~\ref{p.wang}),
but so far has not used implicit quantum test oracles.
The previous section lists properties 
of quantum computing
which might be recast in the form of 
implicit quantum test oracles:

\vspace{1ex}

It would seem that 
it should be straight forward
to code a quantum test oracle
for Section~\ref{sec:pdistribution} 
(probability distributions)
as a sanity check which checks every probability lies in the range 0 to~1.0
and together they sum to~1.0.
However,
perhaps a little care is needed 
to ensure actual calculations are within some epsilon
and choosing an appropriate epsilon.
Such checks should have little overhead and might
be left in the source code even after development.

Similarly checking the width of the quantum circuit
at the output is the same as that at the start
(Section~\ref{sec:numq})
would seem simple and impose little overhead.
Perhaps such sanity checks would be most useful
at the circuit design stage, before the 
circuit is assembled
or the full simulator is run.

Checking that a simulator can indeed be run backwards
(Section~\ref{sec:reversibilty})
would perhaps have too high an overhead to be done
during normal operation.
Indeed a certain care would be needed to choose
both epsilon and the number of forward and backward runs.
For both real circuits and well structured software simulators,
we would expect reversibility to be checked at the base
components (e.g.~quantum gates)
and possibly at intermediate levels
before system tests of the complete circuit and
simulator.
Although testing simulators is always difficult,
Section~\ref{sec:reversibilty} offers 
the gold standard of a known non-trivial system wide result,
which could give confidence in the soundness of 
any quantum simulator.

Unlike our other quantum properties,
entropy conservation
(Section~\ref{sec:entropy})
requires multiple simulation runs.
However it imposes little overhead
and so data could be collected on each run
and entropy conservation checked at a latter state.
Again (Section~\ref{sec:pdistribution})
we are dealing with real number calculations
of probability distributions so
there may be a degree of pragmatism in choosing
the number of runs and epsilon.
Nevertheless conservation of entropy
offers the gold standard reassurance associated with
meeting a known requirement.

\section{Conclusions}

\noindent
Sections~\ref{sec:pdistribution} to~\ref{sec:entropy}
described universal properties of
quantum circuits
and therefore properties that simulators
of quantum circuits
(subject to the fidelity of the simulation)
must also have.

Existing implicit test oracles
(e.g.~null pointer exceptions (NPE),
segmentation errors (segfaults) 
and timeouts)
can be thought of as checks that software is well coded;
the four implicit quantum test oracles
(Section~\ref{sec:discussion})
can be thought of as checking the quantum simulation 
as well as its implementation.
In practise both types should be deployed.

\bibliographystyle{IEEEtran}

\bibliography{/tmp/gp-bibliography,/tmp/references}

\end{document}